\documentclass[a4paper]{article}
\usepackage{amsmath, amssymb, subcaption, cite}
\usepackage{INTERSPEECH2021}

\DeclareMathOperator*{\argmax}{arg\,max}

\newcommand{\refeq}[1]{Eq.~(\ref{eq:#1})}
\newcommand{\reftb}[1]{Table \ref{tb:#1}}
\newcommand{\reffig}[1]{Figure \ref{fig:#1}}
\newcommand{\refsec}[1]{Section \ref{sec:#1}}

\title{Toward Streaming ASR with Non-Autoregressive Insertion-based Model}
\name{Yuya Fujita$^1$, Tianzi Wang$^2$, Shinji Watanabe$^3$, Motoi Omachi$^1$}
\address{
 $^1$Yahoo Japan Corporation, Tokyo, JAPAN\\
 $^2$Johns Hopkins University, MD, USA,  
 $^3$Carnegie Mellon University, PA, USA}
\email{\{yuyfujit, momachi\}@yahoo-corp.jp, wtianzi1@jhu.edu, shinjiw@iee.org}

\begin{document}
\maketitle
\begin{abstract}
Neural end-to-end (E2E) models have become a promising technique to realize practical automatic speech recognition (ASR) systems.
When realizing such a system, one important issue is the segmentation of audio to deal with streaming input or long recording.
After audio segmentation, the ASR model with a small real-time factor (RTF) is preferable because the latency of the system can be faster. 
Recently, E2E ASR based on non-autoregressive models becomes a promising approach since it can decode an $N$-length token sequence with less than $N$ iterations.
We propose a system to concatenate audio segmentation and non-autoregressive ASR to realize high accuracy and low RTF ASR.
As a non-autoregressive ASR, the insertion-based model is used.
In addition, instead of concatenating separated models for segmentation and ASR, we introduce a new architecture that realizes audio segmentation and non-autoregressive ASR by a single neural network.
Experimental results on Japanese and English dataset show that the method achieved a reasonable trade-off between accuracy and RTF compared with baseline autoregressive Transformer and connectionist temporal classification.
\end{abstract}

\section{Introduction}
\label{sec:intro}
End-to-end (E2E) models have become a promising option to realize practical automatic speech recognition (ASR) systems based on the significant improvement of the ASR performance\cite{AttentionNIPS2015, amodei2016deep, LAS2016, Shinji2017hybrid, Chiu2018google, Luscher2019, karita2019comparative, Sainath2019}. 
In particular, the E2E model is suitable for ASR systems run on a device like a smartphone which only has limited computing capability.
Most of the E2E models are composed of a single neural network and its computational optimization is simple. 
Therefore, many techniques to make the model footprint lower like quantization and compression, are easily applied.
This leads to several works that aim to run E2E ASR models on a device like smartphones \cite{He2019, Kim2019, sainath2020streaming}.

One important issue to be solved when realizing a practical ASR system is the segmentation of the input audio. 
If the input audio is long or is fed in a streaming way, the audio should be segmented into an appropriate length of utterances to avoid large memory consumption.
In particular, estimating the end of the speech segment is important because it also affects latency which directly relates to the user's subjective impression of the performance of the ASR system.
However, realizing low latency and high accuracy is a trade-off.
One way to realize such an ASR system is to separately employ models for segmentation and ASR.

In such a scenario, it is obvious that making the decoding process faster leads to lower latency.
Therefore, decoding with a small real-time factor (RTF) is important.
Recently, non-autoregressive Transformer (NAT) is intensively investigated in the research field of neural machine translation\cite{gu2017non, Marjan2019, Gu2019leven, stern2019insertion}. 
It aims to realize faster decoding than an autoregressive Transformer at the expense of a small loss of accuracy. 
The main contribution to the RTF improvement of NAT is the ability to decode $N$-length token sequences with less than $N$ iterations, which can not be achieved by autoregressive Transformer.
Mask-predict, one of the NAT models, was applied to ASR and realized significant improvement of RTF with a small degradation in accuracy\cite{Chen2020Listen}.

In this paper, we aim to realize ASR of streaming input or long recording with non-autoregressive ASR.
Among several works of non-autoregressive ASR\cite{higuchi2020mask, Chen2020Listen, chan2020imputer, fujita2020}, we propose to use the insertion-based model recently proposed in \cite{fujita2020} because of its high accuracy and small numbers of iterations required for decoding.
The model jointly trains connectionist temporal classification (CTC) \cite{Graves2006CTC} and insertion-based models.
It achieves better accuracy with approximately $\log _{\text{2}} (N)$ iterations during inference than autoregressive Transformer with greedy decoding, while the other non-autoregressive models like mask CTC \cite{higuchi2020mask} does not reach this performance.
Our first attempt in this proposed framework is to concatenate audio segmentation and this insertion-based model toward streaming ASR.

In addition, we propose to integrate audio segmentation and non-autoregressive ASR in a single neural network by exploiting the CTC part of the insertion-based model with causal self-attention.
Causal self-attention is realized by block self-attention which is similar to Transformer XL\cite{Dai2019}. It computes attention weights inside a limited context (block). 
In general, introducing causal self-attention degrades accuracy. However, once the audio segment is fixed,  the insertion-based model can refine the hypothesis by looking at the whole acoustic feature in the segment with a smaller number of iterations than the autoregressive Transformer. Therefore, the accuracy can be recovered while achieving faster inference than the autoregressive Transformer.

Experimental results showed the proposed method achieved a good balance between accuracy and RTF compared with baseline autoregressive Transformer and CTC.
To the best of our knowledge, this is the first work to apply non-autoregressive ASR to streaming or long recording.

\section{Related Work}
Several works of non-autoregressive Transformer for ASR have been proposed\cite{higuchi2020mask, Chen2020Listen, chan2020imputer}.
However, all of them do not aim for recognizing streaming input or long recording.
There are some works of realizing audio segmentation and ASR in a single model.
In \cite{Yoshimura2020}, it is proposed to use the CTC part of a jointly trained CTC and attention-based model. 
Integrating an end-pointer into a single RNN-T model and second-stage rescoring with an attention-based model is proposed in \cite{Li2020}. It assumes a voice search application which only detects a single endpoint of streaming audio. 
Both of them uses autoregressive model while ours are non-autoregressive one.

\section{Insertion-based model with CTC}
\label{sec:conv_method}
In this section, the insertion-based model used in this work, KERMIT (Kontextuell Encoder Representations Made by Insertion Transformations)\cite{chan2019kermit}, is first introduced. 
Then, joint modeling with CTC proposed in \cite{fujita2020} is explained.

\subsection{Insertion-based model: KERMIT}
First, the general formulation of E2E ASR models and insertion-based models is described. 
Suppose $X=(\mathbf{x}_t \in \mathbb{R} ^{d} | t=1,\cdots,T)$ is an acoustic feature sequence whose dimension is $d$. 
The output token sequence is defined as $C=(c_n \in \mathcal{V} | n=1,\cdots,N )$.
$T$ is the length of the acoustic feature, $N$ is the length of output token sequence, and $\mathcal{V}$ is a set of distinct tokens. 
The E2E ASR model is a probability distribution over the output token sequence $C$ given the acoustic feature sequence $X$, i.e. $p^{\text{e2e}}(C|X)$. 
It is modeled by a single neural network.

In the case of insertion-based models, the probability distribution $p^{\text{e2e}}(C|X)$ is assumed to be marginalized over insertion order $Z$:
\begin{align}
\label{eq:ins_model}
    p^{\text{e2e}}(C|X) & = \sum _Z p(C, Z|X) = \sum _Z p(C^Z |X) p(Z). 
\end{align}
Insertion order $Z$ represents the permutation of the token sequence $C$, e.g. if $C=(A,B,C,D)$ and $Z=(3,1,2,4)$, $C^Z = (C, A, B, D)$.
For the KERMIT case, let $c^Z_i$ be a token to be inserted and $l^Z_i$ be a position where the token is inserted at the $i$-th generation step under an insertion order $Z$, $p(C^Z |X)$ in \refeq{ins_model} is defined as:
\begin{align}
            p(C^Z|X) 
        =: & \prod _{i=1} ^I p \left( 
            \left( c^Z_i, l^Z_i \right)
                | \left( c^Z_0, l^Z_0 \right), \cdots, \left( c^Z_{i-1}, l^Z_{i-1} \right)
                ,X \right) \nonumber \\
        =  & \prod _{i=1} ^I p\left( \left( c^Z_i, l^Z_i \right) | {c^{Z}}^{\prime}_{0:i-1}, X \right),
        \label{eq:inst_model}
\end{align}
where $i=1,\cdots,I$ is the index of generation step and ${c^{Z}}^{\prime}_{0:i-1}$ is the sorted token sequence at the $(i-1)$-th generation step.
The posterior in \refeq{inst_model} is modeled by only the encoder block of Transformer.
The output at the final layer of the encoder block is sliced as $\mathbf{H}^{\text{tok}} \in \mathbb{R}^{d^{\text{sa}} \times i} $ then used to calculate the posterior:
\begin{align}
      p\left( \left( c^Z_i, l^Z_i \right) | {c^Z}^{\prime}_{0:i-1}, X \right)
    =: \underbrace{p(c^Z_i | l^Z_i, \mathbf{H}^{\text{tok}})}_{\text{Token prediction}} 
        \underbrace{p(l^Z_i | \mathbf{H}^{\text{tok}})}_{\text{Position prediction}},
        \label{eq:inst_pst}
\end{align}
where $d^{\text{sa}}$ is the dimension of self-attention.
The token and position prediction term are calculated by a linear transformation of $\mathbf{H}^{\text{tok}}$ and softmax.
Non-autoregressive parallel decoding is possible using only the token prediction term in \refeq{inst_pst}:
\begin{align}
    \hat{c} = \argmax _{c}  p\left( c|l, {{c}^{Z}}^{\prime}_{0:i-1}, X \right).
\end{align}
When $p(Z)$ is the balanced binary tree (BBT) insertion order proposed in \cite{stern2019insertion}, decoding finishes empirically with $\log_2(N)$ iterations. The BBT order is to insert the centermost tokens of the current hypothesis. For example, given a sequence $C=(c_1, \cdots, c_9)$, the hypothesis grows based on the tree structure like $(c_5) \rightarrow (c_3, c_5, c_7) \rightarrow (c_2, c_3, c_4, c_5, c_6, c_7, c_8) \\ \rightarrow (c_1, c_2, c_3, c_4, c_5, c_6, c_7, c_8, c_9) $.

\begin{figure}[tb]

\begin{minipage}[b]{1.0\linewidth}
  \centering
  \centerline{\includegraphics[width=5.5cm]{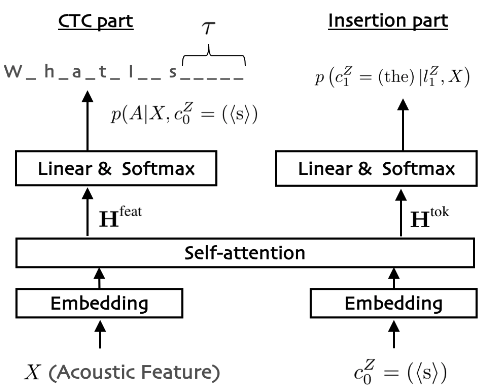}}
\end{minipage}
\caption{Schematic diagram of KERMIT and joint modeling with CTC.}
\label{fig:kermit}
\end{figure}

\begin{figure}[tb]

\begin{minipage}[b]{1.0\linewidth}
  \centering
  \centerline{\includegraphics[width=5.5cm]{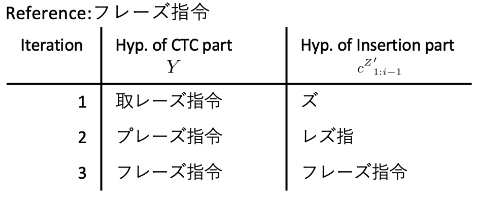}}
\end{minipage}
\caption{Decoding example of KERMIT and CTC.}
\label{fig:decoding}
\vspace{-5mm}
\end{figure}

\subsection{Joint modeling with CTC}
\label{sec:kermit-ctc}
Next, joint modeling of KERMIT and CTC is introduced\cite{fujita2020}. 
In general, joint modeling is to model a joint distribution over different sequences.
Suppose $Y$ is another output token sequence for joint modeling. Then, the posterior in \refeq{inst_model} is modified as:
\begin{align}
    p(C^Z|X)  =:  p(C^Z, Y|X)  
             =  p(Y|X,C^Z) p(C^Z|X).
            \label{eq:joint_ctc}
\end{align}
The term $p(Y|X,C^Z)$ in \refeq{joint_ctc} can be any kind of probability distribution.
We set $Y=C$ and choose CTC for the term because joint modeling with CTC results in faster convergence and performance improvement\cite{Shinji2017hybrid}.
Suppose $A$ is a sequence of tokens including a blank symbol $\langle \text{b} \rangle$, i.e. $A=(a_t \in \mathcal{V} \cup \{\langle \text{b} \rangle \}|t=1,\cdots,T)$. 
$\mathcal{F}(\cdot)$ is a function which deletes repetitions and the blank symbol from the sequence $A$ hence $\mathcal{F}(A) = Y$. 
The CTC probability is formulated as:
\begin{align}
    p(Y|X,C^Z) &=: \sum _{A \in \mathcal{F}^{\text{-1}}(Y)} p(A|X,C^Z).
    \label{eq:ctc}
\end{align}
Because all the output of KERMIT is dependent on both the acoustic feature sequence and the token sequence, the final output of the encoder block of KERMIT is sliced as $\mathbf{H}^{\text{feat}} \in \mathbb{R}^{d^{\text{sa}} \times T}$ and used:
\begin{align}
 \label{eq:kermit_ctc_pst}
    p(A|X,C^Z) &=: p(A|\mathbf{H}^{\text{feat}}).
\end{align}
This process is depicted in \reffig{kermit}.
It can be seen as a multi-task training of the two terms in \refeq{joint_ctc}, i.e. $p(Y|X,C^Z)$ in \refeq{ctc} and $p(C^Z|X)$ in \refeq{inst_model}.
The formulation means CTC is reinforced by insertion-based token sequence generation.
During decoding, either of the CTC part $p(Y|X,C^Z)$ or the insertion part $p(C^Z|X)$ in \refeq{joint_ctc} can be used as a final result.
\reffig{decoding} shows an example of the decoding process. 
The error in the original CTC result ($Y$ in the first iteration) was refined after the iterations.

\begin{figure}[tb]

\begin{minipage}[b]{1.0\linewidth}
  \centering
  \centerline{\includegraphics[width=5.0cm]{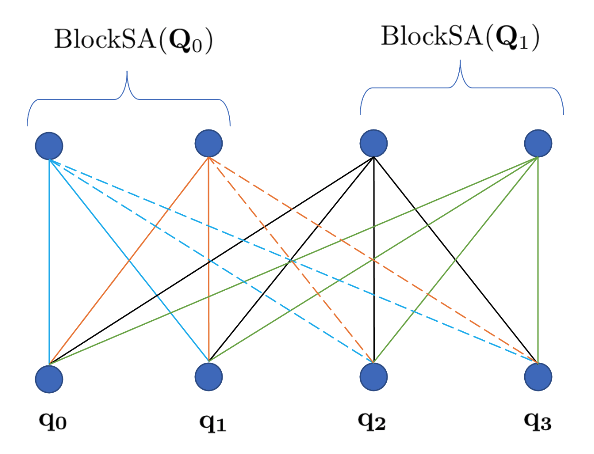}}
\end{minipage}
\caption{Example of the block self-attention ($T_q = 4, L = 2$). In the case of block self-attention, dotted lines are not computed. }
\label{fig:block-selfattention}
\vspace{-5mm}
\end{figure}

\section{Audio segmentation and non-autoregressive decoding}
This section proposes to extend the model described in \refsec{kermit-ctc} to enable joint audio segmentation and non-autoregressive decoding.

\subsection{Preliminary: Block self-attention}
\label{sec:bsa}
In the model described in \refsec{kermit-ctc}, self-attention is the component that prevents the model being used for audio segmentation of long recordings. 
This is because self-attention computes attention weights over the whole input sequence.
In order to avoid this, block self-attention, which computes attention weights inside a limited context (block), is introduced.
This is almost the same structure as Transformer XL \cite{Dai2019} except this formulation uses future context inside a block and passes gradients to the previous block.

First, we define multi-headed attention \cite{TransformerNIPS2017} as follows:
\begin{align}
 \mathbf{U}_h = \text{Attention}(\mathbf{QW}_h^\text{Q}, \mathbf{KW}_h^\text{K}, \mathbf{VW}_h^\text{V}), \\
 \text{MultiHead}(\mathbf{Q},\mathbf{K},\mathbf{V}) = \text{Concat}(\mathbf{U}_1, \cdots, \mathbf{U}_{H})\mathbf{W}^\text{O}, 
\end{align}
where $\mathbf{Q} \in \mathbb{R}^{T^\text{q} \times d^\text{att}}$, $\mathbf{K} \in \mathbb{R}^{T^\text{k} \times d^\text{att}}$, and $\mathbf{V} \in \mathbb{R}^{T^\text{v} \times d^\text{att}}$ denote query, key, and value matrices, respectively. 
$T^\text{q},T^\text{k}$, and $T^\text{v},$ are the length of each elements and $d^\text{att}$ is the dimension of the input to $\text{MultiHead}(\cdot)$. 
$h = 1, \cdots, H$ is the index of the head.
$\mathbf{W}_h^\text{Q}, \mathbf{W}_h^\text{K}, \mathbf{W}_h^\text{V} \in \mathbb{R}^{ d^{\text{att}} \times d^{\text{att}}/H}$ and $\mathbf{W}^\text{O} \in \mathbb{R}^{d^{\text{att}} \times d^{\text{att}}}$. 

Self-attention is the multi-headed attention whose inputs are the same, i.e. $\mathbf{Q}=\mathbf{K}=\mathbf{V}$:
\begin{equation}
 \text{SA}(\mathbf{Q}) = \text{MultiHead}(\mathbf{Q}, \mathbf{Q}, \mathbf{Q}).
\end{equation}
Block self-attention introduces block length $B$ and its index $b$, and define query matrix of $b$-th block $\mathbf{Q}_b \in \mathbb{R}^{B \times d^{\text{att}}}$ as:
\begin{equation}
\label{eq:blocked_query}
 \mathbf{Q}_b = \left[ \mathbf{q}_{b\times B}, \mathbf{q}_{b\times B + 1}, \cdots, \mathbf{q}_{(b+1) \times B -1}\right].
\end{equation}
Then, block self-attention at $b$-th block is defined by setting $\mathbf{Q} = \mathbf{Q}_b, \mathbf{K} = \mathbf{V} = [\mathbf{Q}_{b-1}, \mathbf{Q}_b]$, i.e.:
\begin{align}
    \label{eq:block_sa}
 \text{BlockSA}(\mathbf{Q}_b) = \text{MultiHead}(\mathbf{Q}_b, [\mathbf{Q}_{b-1}, \mathbf{Q}_b], [\mathbf{Q}_{b-1}, \mathbf{Q}_b]).
\end{align}
Note that when computing the $b$-th block, only the $B$ frames of future context are required. This avoids computing attention weights over the whole input sequence, hence realizing segmentation of long or streaming input audio.
Another benefit is computational cost. The self-attention of length $T^{\text{q}}$ requires $\mathcal{O} \left( \left( T^{\text{q}} \right)^2 \right)$ computations. In the block self-attention with block length $B$ requires $\mathcal{O}\left( B^2 \times \lfloor T^{\text{q}}/B \rfloor \right)$.
This computation is depicted in \reffig{block-selfattention}.

\subsection{Proposed joint modeling with block self-attention}
\label{sec:block_kermit}

First, in order to apply joint modeling of KERMIT and CTC in Section \ref{sec:kermit-ctc}, we further extend block self-attention in \refeq{block_sa} to consider extra input $\mathbf{M} \in \mathbb{R}^{T^{\text{m}} \times d^{\text{att}}}$.
This extension is realized by setting $\mathbf{K} = \mathbf{V} = [\mathbf{Q}_{b-1}, \mathbf{Q}_b, \mathbf{M}]$ of multi-headed attention, i.e.:
\begin{align}
    \label{eq:ext_block_sa}
& \text{ExtBlockSA}(\mathbf{Q}_b, \mathbf{M}) \nonumber\\ 
& \quad = 
     \text{MultiHead}(\mathbf{Q}_b, [\mathbf{Q}_{b-1}, \mathbf{Q}_b, \mathbf{M}], [\mathbf{Q}_{b-1}, \mathbf{Q}_b, \mathbf{M}]),
\end{align}
where $T^{\text{m}}$ is the length of extra input.

Now, we consider the acoustic feature sequence $X$.
For ease of explanation, the acoustic feature sequence is assumed to be segmented, e.g. training phase where segment information is given or inference phase where the segment is estimated.
The segmented acoustic feature sequence is passed to an audio embedding layer to obtain an embedding matrix $\mathbf{X}^{\text{E}} \in \mathbb{R}^{{T}^{\text{SS}} \times d^{\text{att}}}$, where $T^{\text{SS}}$ is the segmented sequence length after subsampling.
In order to apply block self-attention, same as \refeq{blocked_query}, $b$-th block of $\mathbf{X}^{\text{E}}$ is defined as $\mathbf{X}_b ^{\text{E}} = \left[\mathbf{x}_{b\times B}, \mathbf{x}_{b\times B + 1}, \cdots, \mathbf{x}_{(b + 1) \times B - 1}  \right]$ where $b = 0, \dots, \lfloor \frac{T^{\text{SS}}}{B} \rfloor - 1$.

In our joint modeling, we use the partial hypothesis $C^{\text{hyp}} = \left( c_{1:i} \right)$ at $i$-th generation step as an extra input, which is artificially generated according to $p(Z)$ in the case of the training phase.
It is passed to a character embedding layer to obtain an embedding matrix $\mathbf{C}^{\text{hyp}} \in \mathbb{R}^{N^{\text{hyp}} \times d^{\text{att}} }$, where $N^{\text{hyp}}$ is the length of the hypothesis.

Thus, the proposed forward pass of KERMIT using block self-attention defined by \refeq{ext_block_sa} is
\begin{align}
 \label{eq:forward}
 \left\{
 \begin{alignedat}{4}
  \mathbf{Z}^{(j)}_b & = &  & \text{ExtBlockSA}(\mathbf{Z}^{(j-1)}_b, \mathbf{Y}_{(j-1)}) ~ \text{for} ~  b = 0, \dots, \lfloor \frac{T^{\text{SS}}}{B} \rfloor - 1,  \\
  \mathbf{Y}^{(j)} & = &  & \text{MultiHead}( \mathbf{Y}^{(j-1)}, \mathbf{K}, \mathbf{V} ) \\
  & & & ~~ \text{where} ~~ \mathbf{K} = \mathbf{V} = \left[ [\mathbf{Z}^{(j-1)}_b]
    _{b=0,\cdots, \lfloor \frac{T^{\text{ss}}}{B}\rfloor -1}, \mathbf{Y}^{(j-1)} \right], \\
 \end{alignedat}
 \right.
\end{align}
where $j=1,\cdots,J$ is the index of the encoder layer and $\mathbf{Z}_b^{(0)} = \mathbf{X}_b^{\text{E}}$ and $\mathbf{Y}^{(0)} = \mathbf{C}^{\text{hyp}}$. Note that $\text{ExtBlockSA}(\cdot)$ and $\text{MultiHead}(\cdot)$ in \refeq{forward} share the parameters.
In the final layer $J$, $\mathbf{Z}^{(J)}$ corresponds to $\mathbf{H}^{\text{feat}}$ in the CTC part of \reffig{kermit} (also introduced in Eq.~\eqref{eq:kermit_ctc_pst}), while $\mathbf{Y}^{(J)}$ corresponds to $\mathbf{H}^{\text{tok}}$ in the insertion part of \reffig{kermit} (also introduced in Eq.~\eqref{eq:inst_pst}).

\subsection{Audio segmentation and decoding}
\label{sec:decoding}
Once the model is trained, audio segmentation and decoding are processed as follows.
Different to the segmented case explained in \refsec{block_kermit}, when the input audio is long or is fed in a streaming way, $b$ in \refeq{forward} is very long or unbounded.
Therefore, we proposed to segment audio by using only the first line of \refeq{forward} because this operation is possible at every $B$ frames of input are obtained by using the embedding $\mathbf{C}^{\text{sos}}$ of $C = \{\langle s \rangle \}$ as $\mathbf{Y}^{(0)}$.
Then, output of CTC probability $p(a_{(b-1) \times B}, \cdots, a_{b\times B}| \mathbf{H}^{\text{feat}} ) $ in \refeq{kermit_ctc_pst} is computed by using $\mathbf{Z}^{(J)}_b$ as $\mathbf{H}^{\text{feat}}$.
If the blank symbol has the highest probability of more than $\tau$ consecutive frames, i.e.:
\begin{align}
    \argmax_ {a_t} p(a_t |\mathbf{H}^{\text{feat}}) = \langle \text{b} \rangle ~ \text{for} ~ t=b\cdot B - \tau, \cdots, b\cdot B, 
\end{align}
the end of the current audio segment is detected as $b \times B$. This is the same strategy proposed in \cite{Yoshimura2020}.
Suppose the index of current audio segment is $r$ and its end as $T_{r}^{\text{end}}$, $r$-th segment is decoded with $b=T_{r-1}^{\text{end}}, \cdots, T_{r}^{\text{end}}$ by both line of \refeq{forward} and iteratively updating $\mathbf{Y}^{(0)}$ by new hypothesis.

\begin{table}[tb]
  \caption{CER and RTF of CSJ when oracle segmentation and full-attention model is used.}
  \vspace{-5mm}
  \label{tb:results_oracle_csj}
 \begin{center}
  \small
    \begin{tabular}{c|ccc|c|}
    & \textit{\textbf{Eval1}} & \textit{\textbf{Eval2}} & \textit{\textbf{Eval3}} & RTF \\
\hline \hline
  ART & \textbf{7.5} & \textbf{5.0} & \textbf{12.2} & 0.98 \\
    CTC & 8.6 & 5.9 & 13.9 & 0.27 \\
 KERMIT & \textbf{7.5} & \textbf{5.0} & \textbf{12.2} & \textbf{0.21} 
    \end{tabular}
    \vspace{-5mm}
 \end{center}
\end{table}

\begin{table}[tb]
  \caption{CER and RTF of CSJ when separated TDNN-based audio segmentation and full-attention model is used. }
  \vspace{-5mm}
  \label{tb:results_tdnn_csj}
 \begin{center}
  \small
    \begin{tabular}{c|ccc|c|}
    &  \textit{\textbf{Eval1}} & \textit{\textbf{Eval2}} & \textit{\textbf{Eval3}} & RTF\\
\hline \hline
  ART &  \textbf{9.8} & \textbf{6.7} & \textbf{14.3} & 0.92  \\
    CTC & 10.5 & 6.8 & 15.3 & 0.31 \\
 KERMIT & 10.1 & \textbf{6.7} & 14.4 & \textbf{0.23} 
    \end{tabular}
    \vspace{-4mm}
 \end{center}
 \vspace{-4mm}
\end{table}

\begin{table}[t]
  \caption{CER and RTF of CSJ when integrated CTC-based audio segmentation with block-SA is used.}
  \vspace{-5mm}
  \label{tb:results_ctc_csj}
 \begin{center}
  \small
    \begin{tabular}{cc|ccc|c|}
    Future &    & & & & \\
    context [sec] &   & \textit{\textbf{Eval1}} & \textit{\textbf{Eval2}} & \textit{\textbf{Eval3}} & RTF\\
\hline \hline
  0.19  & ART & 15.5	& 11.1	& 21.9 & 1.54  \\
   & CTC  & 14.8 & 11.2 & 21.8 & \textbf{0.45} \\
   & KERMIT &  \textbf{11.7} & \textbf{8.6} & \textbf{17.3} & 0.54 \\
   \hline
  0.67 & ART  & 14.1 &9.9 &	19.6 & 1.62  \\
   & CTC  & 12.8 & 9.3 &	18.4 &	\textbf{0.39} \\
   & KERMIT & \textbf{10.8} & \textbf{7.7} &	\textbf{16.2} & 0.45 \\
    \end{tabular}
    \vspace{-4mm}
 \end{center}
\end{table}

\begin{table}[t]
  \caption{WER and RTF of TEDLIUM2. For the integrated CTC-based segmentation, future context is set to 0.19 sec.}
  \vspace{-5mm}
  \label{tb:results_tedlium2}
 \begin{center}
  \small
    \begin{tabular}{c|c|cc|c|}
    Segmentation &  & \textit{\textbf{dev}} & \textit{\textbf{test}} & RTF \\
\hline \hline
     Oracle & ART & \textbf{10.2} &	\textbf{9.1} &	1.11  \\
     & KERMIT & 11.1 &	9.9 &	\textbf{0.19} \\
\hline
    Separated TDNN & ART & \textbf{10.4} & \textbf{13.4} & 0.85  \\
     & KERMIT & 11.8 &	14.2 &	\textbf{0.20}  \\
\hline
    Integrated CTC & ART & 17.9 &	\textbf{19.5} &	1.15  \\
     & KERMIT & \textbf{16.5} &	19.9 &	\textbf{0.38}  
    \end{tabular}
 \end{center}
  \vspace{-6mm}
\end{table}

\section{Experiments}
\subsection{Setup}
The proposed framework is evaluated on the Corpus of Spontaneous Japanese (CSJ)\cite{CSJ2000} and TEDLIUM2\cite{tedlium2}.
Note that for CSJ, only the Academic lecture data, whose amount is around 270 hours, is used for training.
The unsegmented evaluation set is used to evaluate the performance of audio segmentation and ASR.
The segmentation is performed in two ways.
The first one is to employ a separated model for segmentation.
We used a model provided as a baseline of CHiME6 challenge\cite{chime6}. 
The model is based on time-delayed neural network (TDNN) architecture and trained on a target label generated by a highly tuned hybrid ASR model trained on CHiME6 training data.
The second one is using the integrated CTC part as described in \refsec{decoding}.
Baseline models are autoregressive Transformer (ART) and CTC.
Both of the baseline models can use integrated CTC part for segmentation in the same way as described in \refsec{decoding} by using block-SA.
Note that there can be a mismatch of the length of the segmented audio compared with oracle segmentation used in training in both ways. 
Hence it is not apparent that the combination of audio segmentation and non-autoregressive ASR works or not.

Implementation is based on the recipe of ESPnet\cite{espnet2018}.
Most of the parameters are almost the same as used in \cite{fujita2020} except the following two points.
The learning rate and warmup steps were adjusted to stabilize training, and the number of epochs is set to 200 for the KERMIT with block-SA.
The segmentation threshold $\tau$, as discussed in \refsec{decoding}, is set to 8, 16, or 24 depending on the models according to a preliminary experiment.
The RTF is measured with Intel(R) Xeon(R) CPU E5-2640 v4 @ 2.40GHz using two threads for the forward propagation of a neural network.
The beam size of ART and CTC and the number of iteration of KERMIT is set to 5.

\subsection{Results and Discussion}
The character error rate (CER) and real time factor (RTF) of CSJ of each methods are shown in Table~\ref{tb:results_oracle_csj} to \ref{tb:results_ctc_csj}.
The oracle segmentation results in \reftb{results_oracle_csj} showed that KERMIT achieved the best CER, which is the same as ART, but the RTF is smaller than ART. This is the advantage of non-autoregressive ASR.

As shown in \reftb{results_tdnn_csj}, when using the separated TDNN model for segmentation, ART achieved the best CER, but KERMIT also achieved competitive CER with around 1/4 RTF compared to ART.
In the case of using integrated CTC as segmentation, as shown in \reftb{results_ctc_csj}, KERMIT is in a good balance between CER and RTF. 
The RTF of CTC is a bit smaller, but its CER is worse than KERMIT. 
ART was the worst CER and RTF. The integrated approach does not use any alignment nor segmentation criterion while training. This can lead to a significant mismatch between estimated segment length and oracle segment, and the autoregressive model would not be robust on such a mismatch. 
The word error rate (WER) and RTF of TEDLIUM2 are shown in \reftb{results_tedlium2}. Same as the CSJ case, KERMIT is in a good balance between WER and RTF.

In summary, the proposed combination of TDNN segmentation and KERMIT-based NAT achieved a reasonable performance trade-off in terms of RTF and CER.
The performance degradation from the oracle segmentation is acceptable (less than 3\%) while keeping around 0.2 RTF, which is suitable for a streaming scenario.
Another proposal of the integrated CTC approach also shows promising results since it restricts the future context with 0.67 seconds while still keeping 0.45 RTF and within 4\% CER degradation from the oracle segmentation result.
Also, the integrated approach is based on a single neural network, and further optimization can mitigate the issue.


\section{Conclusion}
This paper proposed combining audio segmentation and non-autoregressive ASR toward streaming or long recording audio recognition.
Also, we introduced a new architecture that realizes audio segmentation and non-autoregressive ASR by a single neural network.
The insertion-based model, which is jointly trained with CTC, is used as non-autoregressive ASR. 
By employing causal self-attention, the CTC part is used for audio segmentation. 
Experimental results showed that a combination of audio segmentation and non-autoregressive ASR worked well and achieved a good balance between CER and RTF compared with baseline AR Transformer and CTC.
The single model approach also shows promising results, and the improvement of this approach is left as future work. 

\bibliographystyle{IEEEtran}
\bibliography{mybib}

\end{document}